\newcommand{\PaperTitle}{Enabling Data-Driven Policymaking using Broadband-Plan Querying Tool (BQT+)}
\newcommand{\PaperNumber}{XXX}

\documentclass[10pt,sigconf,letterpaper,nonacm]{acmart}

\usepackage{color}
\usepackage{graphicx}
\usepackage[labelformat=simple]{subcaption}
\usepackage{xspace}
\usepackage{multirow}

\usepackage[english]{babel}
\usepackage{graphicx}
\usepackage{graphics}
\usepackage[utf8]{inputenc}
\usepackage{amsmath}
\usepackage{tabularx}
\usepackage{array, multirow} 
\usepackage{hyperref}
\usepackage{subcaption}
\usepackage{float}
\usepackage[normalem]{ulem}
\usepackage[labelformat=simple]{subcaption}

\newcommand{\ignore}[1]{}

\newcommand{\smartparagraph}[1]{\noindent{\bf #1}\ }

\begin{document}

\title{\PaperTitle}

\author{Laasya Koduru}
\affiliation{%
  \institution{University of California Santa Barbara}
  \city{Santa Barbara}
  \country{USA}
}

\author{Sylee Beltiukov}
\affiliation{%
  \institution{University of California Berkeley}
  \city{Santa Barbara}
  \country{USA}
}

\author{Jaber Daneshamooz}
\affiliation{%
  \institution{University of California Santa Barbara}
  \city{Santa Barbara}
  \country{USA}
}

\author{Eugene Vuong}
\affiliation{%
  \institution{California State University, East Bay}
  \city{Hayward}
  \country{USA}
}

\author{Arpit Gupta}
\affiliation{%
  \institution{University of California Santa Barbara}
  \city{Santa Barbara}
  \country{USA}
}

\author{Elizabeth Belding}
\affiliation{%
  \institution{University of California Santa Barbara}
  \city{Santa Barbara}
  \country{USA}
}

\author{Tejas N. Narechania}
\affiliation{%
  \institution{University of California Berkeley}
  \city{Berkeley}
  \country{USA}
}

\renewcommand{\shortauthors}{Laasya Koduru, Sylee Beltiukov, Jaber Daneshamooz, Eugene Vuong, Arpit Gupta, Elizabeth Belding, and Tejas N. Narechania}

\begin{abstract}
Poor broadband access undermines civic and economic life, a challenge exacerbated by the fact that millions of Americans still lack access to reliable high-speed broadband connectivity. Federal broadband funding initiatives seek to address these gaps, but their success relies on the accuracy of broadband availability and affordability data. This data is often based on self-reported ISP information that overstates coverage and speeds. Such inaccuracies risk misallocating funds, leaving unserved and underserved communities without high-speed Internet. In this work, we present BQT+, an AI-agent data collection platform that queries ISP web interfaces by inputting residential street addresses and extracting the returned data on service availability, quality, and pricing. BQT+ has been applied in policy evaluation studies, including an independent assessment of the state of broadband availability, quality (available speed tiers), and affordability in areas expected to benefit from the \$42.45 billion BEAD program. 


\end{abstract}

\maketitle

\section{Introduction}
Although broadband connectivity is essential, significant disparities in access remain, especially concerning service availability, quality, and affordability. Federal programs seek to bridge these gaps, but their success depends on accurate, granular data at the street address level. Unfortunately, current broadband data is based on unreliable self-reported Internet Service Provider (ISP) information~\cite{nbm_comcast_misreporting, nbm_overstating_coverage}. Independently collected broadband data from ISPs can reveal the true state of broadband service availability, quality, and affordability. Our prior study~\cite{CAF_Sigcomm24} utilizing the Broadband-Plan Querying Tool (BQT) revealed that only 33\% of residential addresses covered by CAF funding can subscribe to broadband meeting the Federal Communications Commission (FCC) service and cost requirements, and only 55\% had access to any service. 

Such findings underscore the need for longitudinal monitoring of federal broadband programs, particularly BEAD.  
Since BEAD-eligible locations are expected to be either unserved ($\leq$ 25/3 Mbps download/upload) or underserved ($\leq$ 100/20 Mbps download/upload), collecting data on broadband service availability (i.e., whether service is offered), quality (what speed tiers are available), and pricing (costs for different tiers to assess affordability) is essential to ensure that (1) federal funds reach communities most in need and (2) the long-term impact of these multi-billion-dollar efforts can be quantified.


\section{Bridging the Data Gaps}
\subsection*{Broadband-Plan Querying Tool (BQT)}
Several prior efforts have sought to independently collect broadband availability and pricing data without relying on ISPs' self-reported data. The California Community Foundation queried Spectrum's ISP interface to curate broadband plans for 165 street addresses in Los Angeles County, and the author in~\cite{tejas_convergence} manually compiled 126 addresses across seven states to collect broadband plans. 
However, these approaches are limited in their ability to effectively evaluate large-scale federal programs.

To facilitate independent collection of broadband availability, quality, and affordability data at scale, our group developed the Broadband-Plan Querying Tool (BQT)~\cite{pauldecoding}, which mimics an end user to automate queries of ISP interfaces. It extracts advertised broadband availability, speed tiers, and pricing data at street-address granularity.  Datasets aggregated by BQT  have supported multiple research studies~\cite{CAF_Sigcomm24,pauldecoding,bst} and policy/legal briefs~\cite{amicus_curiae}, revealing significant gaps between reported and actual service offerings. BQT also offers the potential to enable longitudinal studies of different state/federal policy programs, including BEAD.

\smartparagraph{BQT's limitations.}
The original design of BQT lacked robustness and extensibility, breaking with ISP website updates and requiring intense manual effort to add new ISPs. As BQT is a multi-stakeholder effort, this creates a fundamental tension: democratizing BQT for non-technical stakeholders calls for a no-code interface, yet ensuring extensibility and resilience to website changes demands technical expertise. 

\begin{figure}[hbt!]
   \centering
   
       \includegraphics[width=\linewidth]{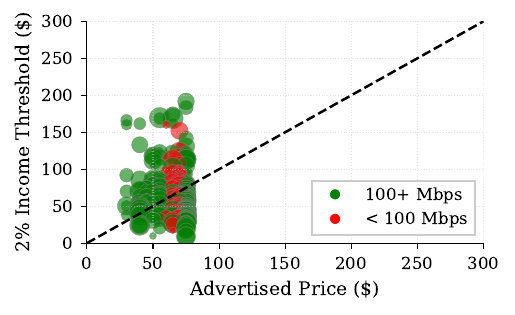}
   \vspace{-0.7em}
   \caption{The state of broadband affordability for BEAD eligible regions in California. 
   Each dot represents a census block group, with green indicating representative plans $\geq$100 Mbps and red indicating representative plans <100 Mbps. Dot size reflects data quality. Dots below the diagonal exceed the 2\% income threshold.}
   \vspace{-0.7em}
   \label{fig:all_states_comparison}
\end{figure}

\vspace{-0.7em}
\subsection*{Proposed Solution: BQT+}
To address the limitations of BQT, we present BQT+, integrated with NetGent ~\cite{daneshamooz2025netgent}, an AI-agent framework, that learns to mimic human interactions with ISP interfaces, balancing robustness and extensibility. BQT+ is designed with agentic capabilities: it interprets high-level instructions, plans actions, and autonomously adapts to interface changes. This balance  allows it to scale across diverse ISPs while reducing manual intervention.

\smartparagraph{Design overview.}
Collecting broadband plan data is challenging since ISP websites vary and often change. While ISP website structures are similar, each ISP's workflow can be mapped as a finite-state machine (FSM) of states and transitions. BQT+ builds an FSM that maps states and transitions by recording a small set of human queries on the ISP's website and synthesizing the observed interactions. BQT+ leverages EasyOCR, an optical character recognition library, to extract all text and their locations from images to identify locations of keywords for each state in the FSM. Once a state is identified, BQT+ employs PyAutoGUI to simulate human mouse and keyboard actions. 

BQT+ separates what a workflow should do from how it is executed. Users provide natural-language state prompts which specify an abstract, non-linear workflow. A State Synthesis component compiles these abstract prompts into concrete states with application-bound detectors and reusable executable code. Concrete states are cached in a repository and deterministically replayed by a State Executor. When an ISP website changes, BQT+ regenerates only the affected states from the same abstract prompts.

This design is extensible, integrating new ISPs with minimal code, and robust, adapting to interface changes with little manual effort. The synthesized FSM is translated into modular code, meeting the no-code requirement for non-technical stakeholders.  BQT+ introduces FSM abstraction, automatic code generation and agentic integration, bridging the gap between accessibility and technical complexity.

\subsection*{BQT+ in Action}
To assess broadband availability and affordability in locations targeted by BEAD, we used BQT+ to systematically collect street-address-level data. Our goal was to evaluate baseline service conditions in BEAD-eligible census block groups (CBGs) and identify potential disparities across four states. 

\begin{table}[t]
  \centering
  \caption{Summary of baseline service conditions, showing the percentage of plans above the 2\% income threshold and below the BEAD speed threshold.}
  \label{tab:summary_findings}
  \resizebox{.95\linewidth}{!}{%
  \begin{tabular}{l|c|c}
    & \textbf{Above 2\% income threshold} & \textbf{Below speed threshold} \\
    \hline
    California & 60\% & 36\% \\
    Michigan & 77\% & 19.3\% \\
    Oklahoma & 74\% & 50\% \\
    Virginia & 61\% & 0\% \\
  \end{tabular}%
  }\\
  \vspace{-.5cm}
\end{table}

BQT+’s design enables scalable collection of broadband data, making it a powerful tool for evaluating large-scale federal interventions such as BEAD. In our case study, its architecture supported 55 ISPs---surpassing the 10 supported through BQT---with ISP integration requiring only a few hours and minimal code.

BQT+ was used to collect broadband data for approximately 63,000 residential addresses in CBGs where at least 50\% of Broadband Serviceable Locations (BSLs) are BEAD-eligible. Within each identified CBG, we randomly sampled 10\% of BSLs in a CBG to query. We focused on four states---California, Michigan, Oklahoma, and Virginia---due to diverse geographical, demographic, and political attributes. Figure~\ref{fig:all_states_comparison} visualizes broadband service availability and quality. Our findings, shown in Table~\ref{tab:summary_findings}, indicate that a significant fraction of BSLs are eligible for federal BEAD funding. Income-based affordability analysis reveals that 65–85\% of representative plan prices exceed the 2\% income benchmark across states. State needs vary---some require affordability interventions, others both infrastructure expansion and affordability measures. These variations underscore the importance of state-specific program design that addresses local market conditions rather than applying uniform national standards across diverse geographic and economic contexts.






\bibliographystyle{ACM-Reference-Format}
\bibliography{refs}

\end{document}